**Częstochowski**

**Kalendarz Astronomiczny**

**2013**

**Rok IX**

**Redakcja**

**Bogdan Wszołek**
**Agnieszka Kuźmicz**

**Wersja elektroniczna kalendarza jest dostępna na stronach**

**www.astronomianova.org**
**www.ptma.ajd.czest.pl**

# Determination of Cycle Length of Quasi-Periodic Signals. Application to Semiregular Variables

Ivan L. Andronov[1], Lidia L. Chinarova[2]

[1] Department "High and Applied Mathematics", Odessa National Maritime University, Ukraine, tt_ari@ukr.net

[2] Astronomical Observatory, Odessa National University, Ukraine

## Introduction - general review

We review methods for determination of "quasi-periods" (or "cycle length") of signals of low coherence. Such type of variability was called "cyclic" for semi-regular red variables, or "quasi-periodic oscillations" (QPO) for fast variability in cataclysmic variables and related objects. These methods may be split into groups of:

the periodogram analysis (Andronov, 1994, 2003)

the wavelet analysis (Andronov, 1998)

the scalegram analysis (Andronov, 1997).

They may be recommended for application for "nearly periodic", "weakly periodic" and "very a-periodic" signals, i.e. for data with decreasing coherence length. The last method is an effective tool for smoothing oscillations with variable shape, "period", phase and mean (averaged over the cycle) value and is independent on linear and even parabolic trends. For the flickering, it shows a "fractal-type" power law dependence of the unbiased estimate of the r.m.s. deviation of the signal from the fit $\sigma$ on the filter half-width $\Delta t$: $\sigma \sim (\Delta t)^\gamma$, where the parameter $\gamma = \gamma_\sigma = 0.5$-$D$, and $D$ is a fractal dimension (Andronov et al., 1997).

For the QPO, the so-called "$\Lambda$-scalegram" was introduced as an extension of the "$\sigma$-scalegram" proposed earlier, which allows to determine effective values of the period and semi-amplitude, as well as an additional parameter related to coherence.

This method was applied to 173 semi-regular variables. Results were compiled in the catalogue. This method is more effective than that of the periodogram analysis, if the signal is of low coherence. It may be also more effective that the wavelet analysis, if the signal undergoes significant low-frequency trends. Some discussion may be found in Andronov and Chinarova (2003).

The parameters are studied in the connection to subtype of variability (SRa, SRb, SRc, SRd) and may be used for an additional classification of long-period late-type pulsating variables.

## Application to AF Cyg

As an illustration, we present a study of the SRb-type pulsating variable AF Cyg based on 8738 observations from the AFOEV database for a recent interval JD 2451626-2455378. Other "unsure" and "fainter than" data were removed before the time series analysis. Study of previous photometric behavior was presented by many authors. The variability was discovered by Espin (1898). Kanda (1922) noticed a large inequality in the period of AF Cygni, and provided new elements assuming a sinusoidal term in the O-C and mentioned period variations from $79.4^d$ to $97.4^d$ with a secondary long-term period of $4300^d$ and a mean period of $88.4^d$. Vorontsov-Velyaminov (1925) mentioned, that the variations of O-C from a similar period of $88.59^d$ are not sinusoidal and calculated parameters of harmonics of this main period, the amplitudes of which do not decay rapidly, as assumed for smooth periodic variations. O'Connell (1932) based on much more observations, determined a period of $94.1^d$ and mentioned that "it is not easily to assign a mean period to a star that varies as capriciously as AF Cygni appears to do" and that " with such large changes in range and shape of light curve, a mean light curve is out of question". The range of estimates of the period is $80.2^d$ -$105^d$.

Kopal (1933) suggested that similar stars should be a base of a separate "AF Cyg-type" class and should be at an evolutionary stage between long-periodic variables and the RV Tau-type stars. He proposed to double the period, assuming two unequal minima similar to RV-type stars. In this case, the "double" period varies from $182.4^d$ and $190^d$. Klius (1983) made an auto-correlation analysis and reported on "independent" brightness oscillations with mean periods of $93^d$, $176^d$, and $941^d$, however, mentioning "the 93 day and 176 day cycles to predominate alternately".

Andronov and Chernyshova (1989), based on the O-C analysis, detected the switches between two shorter periods, which may correspond to changes between the pulsation modes. The "lifetime" of each pulsation mode may last from few to few dozens cycles. For detailed review on evolutionary status of long-period variables, see e.g. a review by Kudashkina (2003).

Kiss et al. (1998) reported on 3 periods in AF Cyg of $921^d$, $163^d$ and $93^d$ with corresponding amplitudes of $0.08^m$, $0.11^m$ and $0.11^m$, respectively. These values of "periods" are close to that reported by Klyus (1983) and Andronov and Chernyshova (1989) based on other methods, but these previous results are not referred.

Using the wavelet analysis for irregularly spaced data (Andronov, 1998), the optimal wavelet smoothing was applied to AF Cyg (Andronov, 1999) with taking into account dependence of the statistically optimal period on time (which has a character of abrupt switches).

The part of the light curve for the time interval mentioned above is shown in Fig.1. The periodogram analysis using sine fit (Andronov, 1994, 2003) was used for a preliminary period determination, with a subsequent correction using a statistically optimal trigonometric polynomial fit of order $s$. The periodogram is shown in Fig.2. For these data, $s=1$, i.e. no significant harmonics are detected. The range of smoothed brightness variations is from $7.043^m \pm 0.006^m$ to $7.522^m \pm 0.006^m$, i.e. the total amplitude is $0.478^m \pm 0.006^m$ (here we take into account correlations between errors of parameters). Individual observations range from $6.1^m$ to $8.3^m$. The photometric elements are

$$\text{Max.JD}=2453260.2\ (\pm 0.3)+ 94.187(\pm 0.026)E.$$

One may note that the fit does not follow all the cycles. Sometimes (at the beginning) the fit and individual cycles of pulsations are out of phase, thus the amplitude is small.

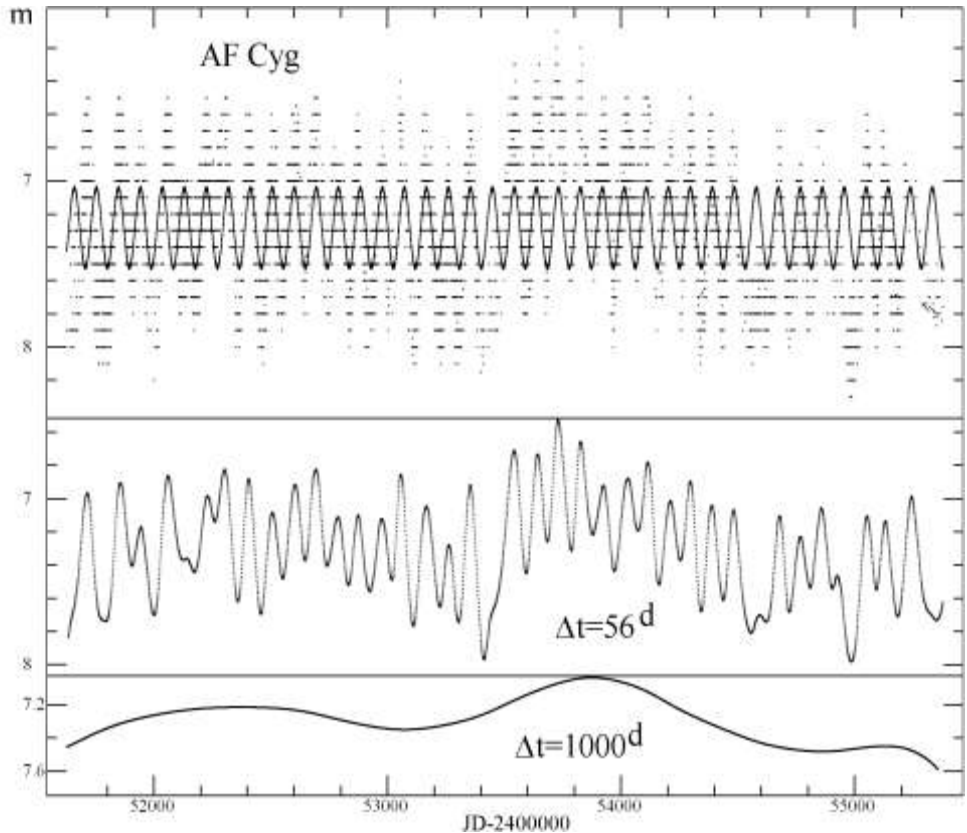

Fig.1. Original observations (points), best sinusoidal fit and "running parabola" fits for different values of the filter half-width $\Delta t=56^d$ and $\Delta t=1000^d$, which correspond to maxima of the "signal-to-noise" ratio in Fig.3.

In this work, we apply the method of "running parabolae" with an additional local weight function $p(z)=(1-z^2)^2$, if $-1<z<+1$ where $z=(t-t_0)/\Delta t$, $t_0$ is trial time and $\Delta t$ is filter half-width (in the wavelet terminology, "shift" and "scale", respectively). Details were presented by Andronov (1997). In Fig.1, we show original observations, and fits corresponding to local maxima of the "signal-to-noise" ratio at $\Delta t=56^d$ (S/N=13.1) and $\Delta t=1000^d$ (S/N=12.9). With an increasing $\Delta t$, the r.m.s. amplitude of the fit decreases, but also decrease an accuracy estimate of the fit. Thus one may choose different $\Delta t$ to study variability at different time scales. The position of the maximum of the test function $\Lambda(\Delta t)$ may allow to estimate the "effective period" (cycle length), whereas its height determines the effective amplitude. This method is especially suitable for noisy signals of low coherence (cyclicity rather than periodicity).

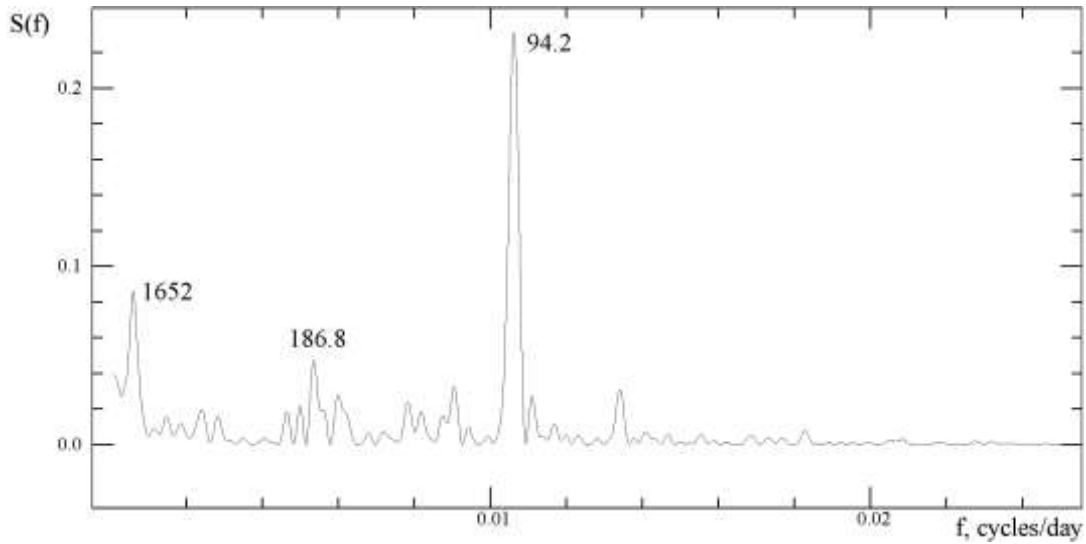

Fig.2. Periodogram S(f) for AF Cyg. Three highest peaks are marked with values of corresponding periods.

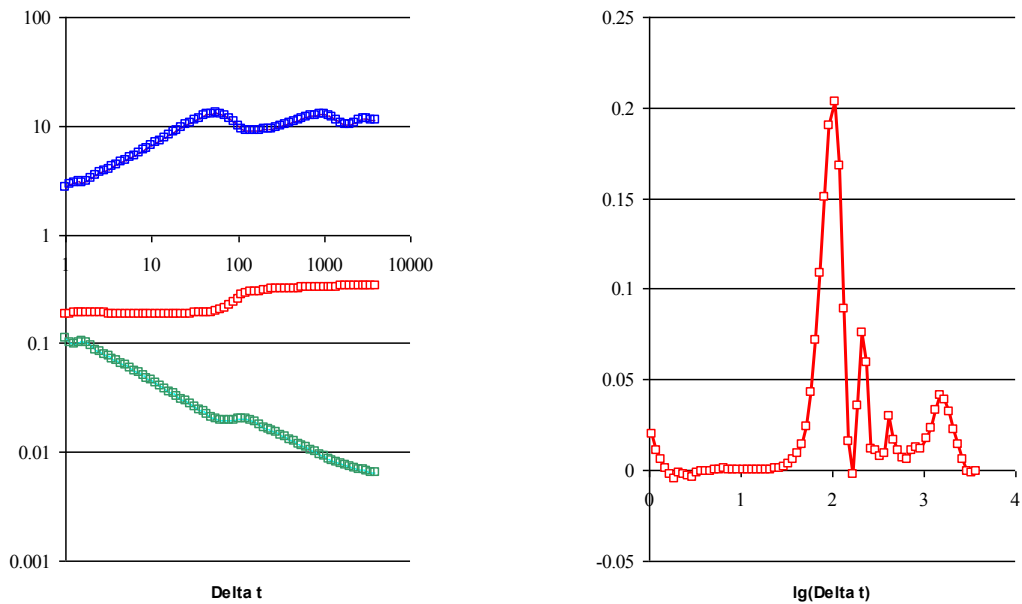

Fig.3. Scalegrams using "Running parabola": left: "signal-to-noise" ratio (up), unbiased estimate of r.m.s. deviation of the observations from the fit, i.e. "σ-scalegram, Andronov (1997)) (middle), r.m.s. accuracy estimate of the fit at moments of observations (down); right: "Λ-scalegram" (Andronov 2003). Three peaks at $\Lambda(\Delta t)$ correspond to effective periods of $93.2^d$, $195.7^d$, $1390^d$ and semi-amplitudes $0.284^m$, $0.175^m$, $0.128^m$, which agree with corresponding estimates from the periodogram analysis, but do not suggest strict periodicity during an interval of observations.

## References

Andronov I.L., 1994, Odessa Astron. Publ., 7, 49
Andronov I. L., 1997, A&A Suppl., 125, 207
Andronov I.L. et al., 1997, Odessa Astron. Publ., 10, 15
Andronov I. L., 1998, Kinem. Phys. Celest. Bodies, 14, 374
Andronov I. L., 1999, in: Self-Similar Systems, ed. by V.B. Priezzhev and V.P.

Spiridonov (JINR, E5-99-38, Dubna), 57, http://uavso.pochta.ru/dubna.pdf
Andronov I. L., 2003, Astron. Soc. Pacif. Conf. Ser., 292, 391
Andronov, I. L., Chernyshova, I. V., 1989, Astron. Tsirk., 1538, 18
Andronov I.L., Chinarova L.L., 2003, Astron. Soc. Pacif. Conf. Ser., 292, 401
Espin, 1898, Astron. Nachr. 145, 327
Kanda S., 1922, Japan. J. Astron. Geophys. 1, 211
Klyus I. A., 1983, Astron. Zhurn. 60, 91
Kopal Z.: 1933, Astron. Nachr., 250, 15
Kudashkina L.S., 2003, Kinem. Phys. Celest. Bodies, 19, 193
O'Connell D. J. K., 1932, Harvard College Obs. Bull. 888, 1
Vorontsov-Velyaminov B., 1925, Astron. Nachr. 225, 369

\*\*\*

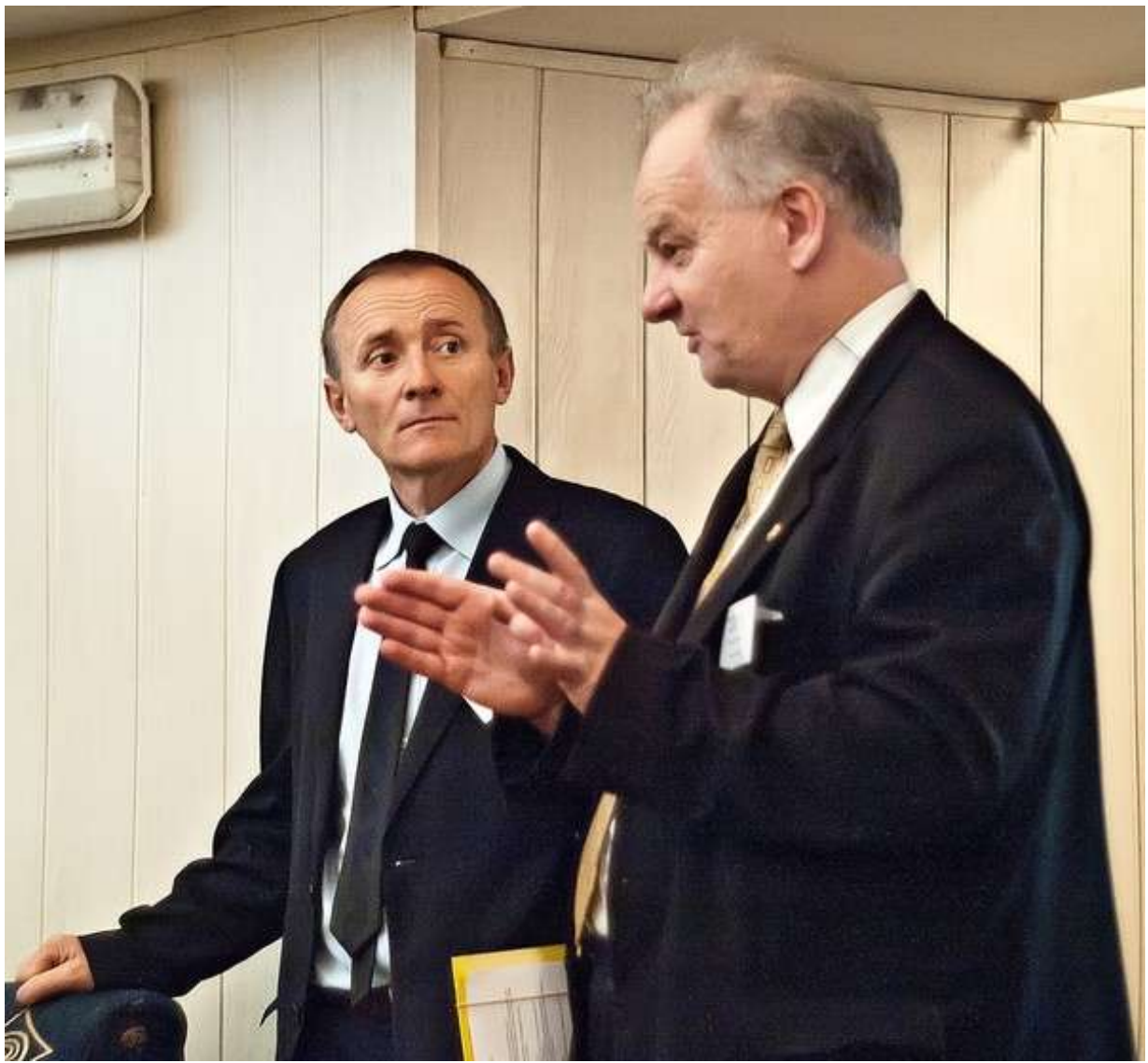

Ivan L. Andronov i Bogdan Wszołek. (*archiwa Astronomii Novej*)